\documentclass[conference]{IEEEtran}
\IEEEoverridecommandlockouts
\usepackage{cite}
\usepackage{amsmath,amssymb,amsfonts}
\usepackage{algorithmic}
\usepackage{graphicx}
\usepackage{textcomp}
\usepackage{xcolor}
\def\BibTeX{{\rm B\kern-.05em{\sc i\kern-.025em b}\kern-.08em
    T\kern-.1667em\lower.7ex\hbox{E}\kern-.125emX}}
\begin{document}

\title{An investigation into the performances of the Current state-of-the-art Naive Bayes, Non-Bayesian and Deep Learning Based Classifier for Phishing Detection: A Survey
}

\author{\IEEEauthorblockN{1\textsuperscript{st} Tosin Ige}
\IEEEauthorblockA{\textit{Dept. of Computer Science } \\
\textit{The University of Texas at El Paso}\\
Texas, USA \\
toige@miners.utep.edu}
\and
\IEEEauthorblockN{2\textsuperscript{nd} Christopher Kiekintveld}
\IEEEauthorblockA{\textit{Dept. of Computer Science} \\
\textit{The University of Texas at El Paso}\\
Texas, USA \\
cdkiekintveld@utep.edu}
\and
\IEEEauthorblockN{3\textsuperscript{rd} Aritran Piplai}
\IEEEauthorblockA{\textit{Dept. of Computer Science} \\
\textit{The University of Texas at El Paso}\\
Texas, USA \\
apiplai@utep.edu}
\and
\IEEEauthorblockN{4\textsuperscript{th} Amy Wagler}
\IEEEauthorblockA{\textit{Dept. of Public Health Science} \\
\textit{The University of Texas at El Paso}\\
Texas, USA \\
awagler2@utep.edu}
\and
\IEEEauthorblockN{5\textsuperscript{th} Olukunle Kolade}
\IEEEauthorblockA{\textit{Office of Naval Research} \\
\textit{United State Navy}\\
Pentagon, USA \\
ookol@unc.edu}
\and
\IEEEauthorblockN{6\textsuperscript{th} Bolanle Hafiz Matti}
\IEEEauthorblockA{\textit{Office of Network Security} \\
\textit{Palo Alto Networks Inc}\\
Texas, USA \\
bolanle.matti@gmail.com}
}

\maketitle

\begin{abstract}
Phishing is one of the most effective ways in which cybercriminals get sensitive details such as credentials for online banking, digital wallets, state secrets, and many more from potential victims. They do this by spamming users with malicious URLs with the sole purpose of tricking them into divulging sensitive information which is later used for various cybercrimes. In this research, we did a comprehensive review of current state-of-the-art machine learning and deep learning phishing detection techniques to expose their vulnerabilities and future research direction. For better analysis and observation, we split machine learning techniques into Bayesian, non-Bayesian, and deep learning. We reviewed the most recent advances in Bayesian and non-Bayesian-based classifiers before exploiting their corresponding weaknesses to indicate future research direction. While exploiting weaknesses in both Bayesian and non-Bayesian classifiers, we also compared each performance with a deep learning classifier. For a proper review of deep learning-based classifiers, we looked at Recurrent Neural Networks (RNN), Convolutional Neural Networks (CNN), and Long Short Term Memory Networks (LSTMs). We did an empirical analysis to evaluate the performance of each classifier along with many of the proposed state-of-the-art anti-phishing techniques to identify future research directions, we also made a series of proposals on how the performance of the under-performing algorithm can improved in addition to a two-stage prediction model
\end{abstract}


\section{Introduction}
Phishing is a type of cybercrime in which an individual is lured to divulging sensitive information details through text message, email, or phone conversation by someone posing either as a legitimate institution or a member of a legitimate institution, some of these commonly requested sensitive details which are social security number, password, credit, and banking card details etc are later used to access more sensitive information for a different type of cybercrime which often results in financial loss or identity theft as about 76\% of the phishing attacks were credential-harvesting in 2022 according to Digital Information world. A California teenager was able to get sensitive information to access credit card details and withdraw money from his victim's account through his fake "America Online" website which resulted in the first lawsuit filed in 2004. Efficient phishing detection has been challenging as attackers continue to advance their tactics as technologies evolve. To defraud personnel, all an attacker needs to do is simply clone a legitimate website to create a new website (SCAM Website) which is then used to defraud computer users.

Email phishing is responsible for 90\% of ransomware attacks and for which the average ransom payment in those instances is can be as high as \$200,000 (£161,000), and in addition to the fact that organizations that fall victim of ransomware attacks lose a couple of weeks as downtime \cite{moller2023ransomware}. The UK Government’s Cyber Security Breaches Survey of 2022 had revealed that cyberattacks rose by 38\% in 2022 alone compared to 2021 as 83\% of businesses and organizations have suffered at least one data breach with Over 3.4 billion phishing emails sent daily. According to the U.S Federal Bureau of Investigation, 2 billion dollars were stolen due to phishing in 2018 alone, 5 billion dollars was stolen in 2019, and 4.7 billion in 2021 \cite{rajeswary2023comprehensive,okomayin2023data,adewale2023encoder}.In 2019, insights Business E-mail Compromise (BEC) announced that about 4.8 million dollars were lost as a result of phishing attacks in 2022, while a cybersecurity research group reported that a whopping 1.6 million dollars were lost in 2019 4.7 billion dollars in 2021 during covid-19 pandemic due to phishing attack.

\begin{figure}
\centering
\includegraphics[width=1 \linewidth]{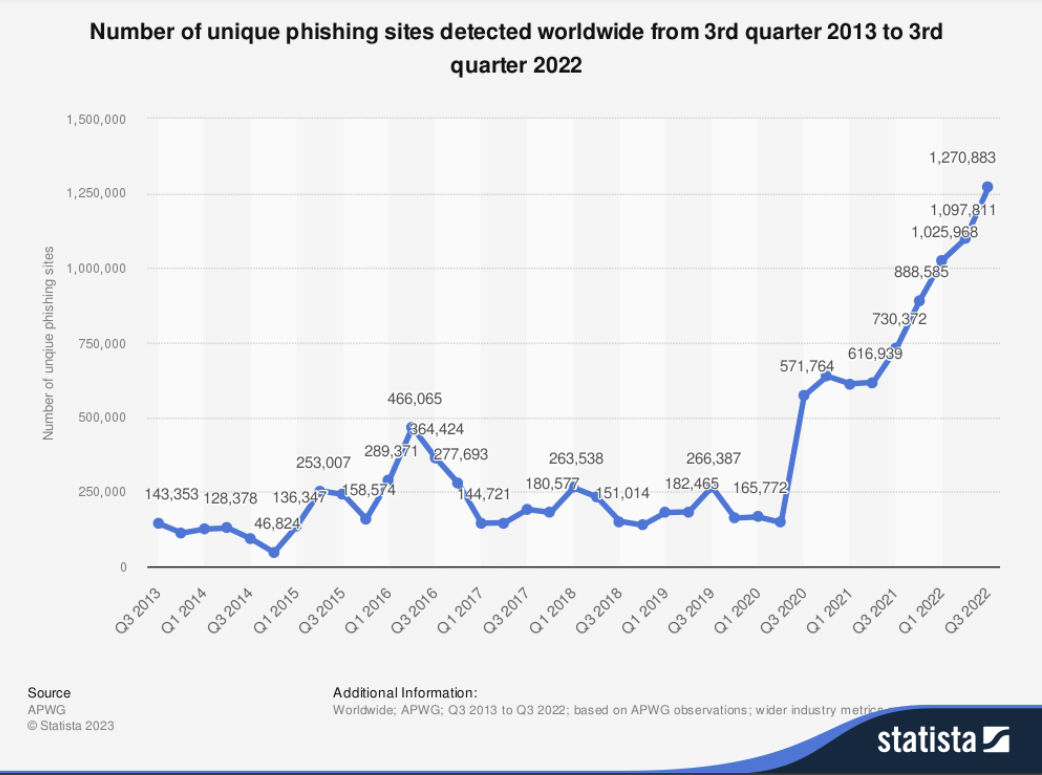}
\caption{\label{fig:Phishing}Phishing statistics from 2013 Q3 to 2022 Q3.}
\end{figure}

The ever-evolving ways attacker tries to improve their phishing techniques to bypass existing state-of-the-art anti-phishing detection and prevention method poses a mountain of challenge to researchers in both industry and academia. Thus, the constant evolvement and innovation in phishing techniques adopted by attackers are the reason why all existing anti-phishing methods remain vulnerable to phishing attacks. All existing methods of detecting phishing attack which are based on machine learning \cite{abdulrahman2023web,aljofey2022effective,anitha2022new,ige2022implementation,ige2022enhancing,ige2023adversarial,ige2024deep,ige2024investigation}, blacklists/whitelists \cite{ghaleb2022phishing}, natural language processing \cite{jain2019machine}, visual similarity \cite{jain2019machine}, rules \cite{jain2016novel},  remains vulnerable to attack due to the following reasons;

\begin{itemize}
\item Very small or minute changes to the uniform Resource Locator (URL) of a blacklisted URL will make the blacklist/ whitelist phishing detection method to fail. Also, the fact that there is no worldwide centralize database  for whitelisted or blacklisted URL make this method even more vulnerable, and so if company X blacklisted my phishing URL on their internal server, I can try it with company Y and be successful.
\item In machine learning phishing detection which uses relevant features such as URL, webpage content, website traffic, search engine, WHOIS record, and Page Rank has their own vulnerabilities because firstly, such classifier will make a phishing URL that is hosted on a hacked or compromise server to be false classify as benign leading to false negative, secondly using domain age as a feature to train a model will always lead to higher false positive simply because the URL of a newly registered legitimate company website will be misclassify because the domain name was recently register, page rank is zero, and with low traffic, and thirdly the fact that parameters for those features are gotten from third party website is another concern. What will happen if the third party website is having a downtime?
\item The issue with visual similarity-based heuristic method which compares both the pre-stored signature such as images, font styles, page layout, and screenshot and so on of the new website with the old website will have general difficulty in detecting anomaly in a newly hosted phishing site.
\item The fact that the majority of the existing machine learning models are trained based on textual features such as “\#”,”.”, Internet Protocol address, URL Length, domain levels, and so on from the Uniform Resource Locator (URL) does not help simply because any phisher or attacker with little web technologies can develop what we called "friendly URL” depending on the programming language adopted whether JAVA, C\#, Python, PHP or framework to avoid all those features. With a friendly URL, such models are bound to misclassify leading to an increment in false negative rate.
\end{itemize}

For any Machine learning-based phishing detection method to be effective in real-time combat against phishing attacks, it must address each of the stated reasons above for which existing state-of-the-art anti-phishing techniques continue to be vulnerable as phishing methods continue to evolve in a more sophisticated and innovative way. It is worth noting that past reviews on phishing have been largely based on approaches, classification, and so on. RASHA ZIENI et al. \cite{zieni2023phishing} focus their review on list-based, similarity-based, and machine learning-based categories of approaches for phishing detection to identify pending research gap, Angad et al. \cite{muneer2021survey} focus theirs on the advantages and limitations of existing approaches to phishing detection, while also using discussion of related application scenarios as guidance to propose a new method of anti-phishing detection, Yifei Wang \cite{wang2022survey}  categorizes widely used phishing detection methods into seven categories and summarizes them.

In this work, we did an extensive review of some of the most recent works on phishing detection, and state-of-the-art algorithms from the past 5 years in order to investigate the performance of the Naive Bayes algorithm relative to other state-of-the-art algorithms for phishing detection task, and the factors behind those performances to uncover future research direction. Our first strategy was to Isolate Naive Bayes from other algorithms, hence, we categorized state-of-the-art phishing detection classifiers into Naive Bayes-based, Machine learning-based, and Deep learning-based for better analysis. The contributions of our research are stated below;

\begin{enumerate}
\item Comparative study of the performance of Naive Bayes relative to other machine learning and deep learning-based state-of-the-art algorithms for phishing detection tasks through a survey of the recently published research works.
\item Investigating and analyzing possible factors behind our findings on the performance of Naive Bayes relative to other machine learning and deep learning-based state-of-the-art algorithms for phishing detection
\item Proposing  possible solutions so as to identify future research direction

\end{enumerate}

The rest of the paper is organized as follows:
Section II some of the most common forms of phishing attacks by which several high-profile attacks have been carried out in recent years, Section III is split into 3 subsections of Bayesian-based, Non-Bayesian-based, and Deep Learning-based based on the categories of state-of-the-art phishing detection we are considering, under each subsection, we described existing state-of-the-art algorithm under their category. In section IV, we looked at the current approaches for phishing detection, this section is further divided into two subsections based on the two major categories of phishing detection approaches, so, under each section, we described different phishing detection techniques under each subsection, we also looked at the limitations of the current state of art phishing detection methods, while our findings were analyzed and discussed in section V. Finally, conclusion and possible future research directions based on our findings were presented in section VI. 

\section{BACKGROUND STUDY}
Since it is easier for attackers to exploit human weakness to easily bypass the most advanced state-of-the-art defense system by extracting sensitive credentials and information through phishing. Attackers therefore focused their effort on getting sensitive credentials through phishing emails which are mistaken for legitimate emails by unsuspecting victims. Hence, it is imperative to understand how different phishing technique works in order to proffer a strategic defense solution to effectively detect, prevent or mitigate phishing impact in case of a successful attack. In this section, we analyse the process of the major phishing attack.
\subsection{Email Phishing}
Email phishing is a phishing type in which unsuspecting victim is tricked into divulging credential or sensitive information through email \cite{aljofey2022effective, li2019stacking}. Here the attacker sends phishing code either through email containing a phishing link or malware attachment in such a way that as soon as the victim clicks on the link \cite{burita2021analysis}, it will either redirect it to a phishing site or get the system infected by malware. Sensitive credentials getting by this mean can then be use by the attacker to commit series of cybercrimes against the victim or target organization including but not limited to remote malware installation, instigate Denial of service attack, Cyberstalking, identity theft, and can even be sold in the dark market.
\subsection{Spear Phishing}
Statistic from Barracuda data shows that a typical organization receives 5 customized spear phishing email each day targeting an individual, and despite the fact that only 0.1\% of all emails are spear phishing attacks, 66\% of all organization breaches are caused by spear phishing. In this type of attack, the attacker keep tracks of the prospective victim activities \cite{sahingoz2019machine,haruta2017visual} in the social media such as X formally Twitter, Linkedin, Facebook, Instagram and so on so as to gather substantial information about the targeted victim. With this newly gathered information, the attacker is able to compose email messages which will seems to come from the organization's manager account and typically requesting for sensitive information belonging to the organization.
\subsection{Voice Phishing (Vishing)}
It is a type of cybercrime in which attacker make automated phone call by a seemingly legitimate phone number from an organization to get confidential detail from unsuspecting victim \cite{cook2008phishwish}. An instance is a customer who get a warning call from an attacker who posed to be bank staff claiming u usual activities on the victim's account and requesting for recently generated one-time password (OTP) or Personal Identification Number (PIN) of the account. The fact that the phisher was able to make scam call from an organization which the victim has connection with makes gives this type of attack a high success rate as experienced in 2021 when 59.49 million which is a whopping 23\% of the America population lost an estimated 29.8 billion US Dollar to voice phishing according to earthweb.
\subsection{SMS Phishing (Smishing)}
It is a type of cybercrime in which a bait message is sent by an attacker to a set of targeted audience through text message. Messages in a smishing attack usually contains either an email to contact, phone number to call, or link to click where the potential victim is then to provide person credential information such as credit card details, password etc for later use by the attacker on legitimate website to commit series of cybercrime. The SMS uses series of social engineering tactics to ensure potential victim follow the instruction by calling the phone number, contacting the email, or clicking on the link which will lead to the actual phishing website with a form to collect their personal data.

\section{Category of Current State of the art Phishing Detection model}
\subsection{Bayesian-Based-Classifier}
Naive Bayes is a family of probabilistic-based algorithms that is based on the Bayes rule. It is based on the fact that, if B has occurred, we can find the probability that A will occur. B is taken to be the evidence while the hypothesis is A and with a strong assumption that each of the features is independent. It uses the prior probability distribution to predict the posterior probability of a sample that belongs to a class. In this process, the class with the highest probability is then selected as the final predicted class \cite{wang2023combination}. Naive Bayes updates prior belief of an event occurring given that there is new information. Hence, given the availability of new data, the probability of the selected sample occurring is given by;
        
        \begin{align*}
            P(class/features) = \frac{P(class) * P(features/class)}{P(features)}
        \end{align*}

\hspace{0.1cm}Where
        
        \begin{itemize}
            \item P(class/features) : Posterior Probability
            \item P(class) : Class Prior Probability
            \item P(features/class) : Likelihood
            \item P(features) : Predictor Prior Probability
        \end{itemize}

It has a very strong assumption of independency which affects its performance for classification tasks\cite {ige2023performance} as the strong assumption of independence among features is not always valid in most of the dataset that is used to train the current state-of-the-art model for several classification tasks. The strong assumption of the Naive Bayes classifier is one reason why it usually underperforms when compared with its peers for similar classification tasks. Naive Bayes classifier has different variants with each variant having its own individual assumption which also impacts its performance in addition to the general assumption of independence which is common to all variants of the Naive Bayes classifier, and so each variant is suitable for different classification tasks.

Multinomial Naive Bayes is a variant of Naive Bayes, It assumes multinomial distribution among features of dataset in addition to the general assumption of independency, and so its performance is affected if the actual distribution is not multinomial or partially multinomial. Multinomial Naive Bayes is the suitable variant for natural language processing classification task \cite{ige2022ai} but still underperforms when compared with non-bayesian and deep learning-based classifiers for the same NLP classification task.

Gaussian Naive Bayes is the suitable Bayesian variant for anomaly detection in network intrusion which could be used to detect Distributed Denial of Service (DDOS) attacks \cite{ige2023performance}. It assumes the normal distribution among features in dataset in addition to the general assumption of independence which is common to all variants of Naive Bayes. 


Despite being a suitable Naive Bayes variant for anomaly detection, it still underperforms when compared with its suitable peer for detection of Distributed Denial of Service (DDOS) attack as evident in the work done by Rajendran \cite{rajendran2023improved} where Gaussian Naive Bayes have the least accuracy of 78.75\% compared with other non-bayesian based for attack detection classification task.

Bernoulli Naive Bayes assumes Bernoulli distribution in addition to the assumption of independence. Its main feature is that it only accepts binary values such as success or failure, true or false, and yes or no as input while complement Naive Bayes is used for imbalance datasets as no single variant of Naive Bayes can do the task of all the variants. Both the suitability and performance of each variant are determined by their individual assumption in addition to the general assumption of independence which impacts their performance when compared with their suitable peer for the same classification task.

\subsection{Non-Bayesian Based Classifier}
\subsubsection{Decision Tree}
A decision Tree is a Supervised learning technique whose operation is based on a tree-structured classifier, with features in the dataset being represented by an internal node, each decision rule is represented by the branches, while the internal nodes represent the features of a dataset, branches represent the decision rules and each leaf node represents the decision outcome is represented by the leaf node and so does not have further branches. It makes a decision-based graphical representation of all possible solutions to a problem. It uses the Classification and Regression Tree algorithm (CART) \cite{zhu2020dtof} to construct a decision tree starting with the root node whose branch keeps expanding further to construct a tree-like structure. It is a non-parametric and the ultimate goal is the creation of a machine learning model capable of making prediction by learning simple decision rules that are inferred from data features.

\subsubsection{Random Forest}
It is an ensemble-based learning algorithm that could be used for classification, regression task, and other similar tasks that operates based on the construction of multiple decision trees \cite{hr2020development}. Since the algorithm works by constructing multiple decision trees during training, the output of a classification model trained with a random forest algorithm is the class selected by most of the trees, while the mean or average prediction of individual trees is returned as the output for a regression task. This system of aggregating and ensemblement with multiple trees for prediction makes it possible for a random forest-trained model to outperform the decision tree-trained model and also avoid overfitting which is a peculiar problem for decision tree classifiers.

\subsubsection{Logistic Regression}
Logistic regression is the modeling of the probability of a discrete outcome by having the event log-odds be a linear combination of one or more independent variables given an input variable \cite{EDGAR201795}. Logit transformation is applied to the bounded odds which is the division between the probability of success and probability of failure based on
it's linear regression that could be used for both classification and regression tasks and since the output is a probability, the dependent variable is bounded between 0 and 1 values, it uses logistic function to model binary output for classification problems. The difference between linear regression and logistic regression is that the range in logistic regression is bounded by 0 and 1, and also that logistic regression does not require a linear relationship between input and output.

\subsubsection{XGBoost}
It is a supervised learning algorithm that is gradient boosting based. It is extremely efficient and highly scalable, the algorithm works by first creating a series of individual machine learning models and then combining each of the previously created models to form an overall model that is more accurate and efficient than any of the previously created individual models in the series. This system of creating a series of models and combining them to create a single model \cite{gu2022ensemble} makes XGBoost perform better than other state-of-the-art machine learning algorithms in many classification, ranking, several user-defined prediction problems, and regression tasks across several domains. XGboost uses gradient descent to add additional individual models to the main model for prediction, hence it is also known as stochastic gradient boosting, gradient boosting machines, or multiple additive regression trees.

\subsubsection{K-Nearest Neighbor (KNN)}

k-nearest neighbors (kNN) algorithm is a non-parametric supervised learning algorithm that uses the principle of similarity to predict the label or value of a new data point by considering values of its K-nearest neighbors in the training dataset based on a distance metric like  Euclidean distance. 
\begin{equation}
\textrm{dist}(x,z) \leq \textrm{dist}(x,y) + \textrm{dist}(y,z)
\end{equation}

for which the distance between x and z could be calculated by

\begin{equation}
 d\left( x,z\right)   = \sqrt {\sum _{i=1}^{n}  \left( x_{i}-z_{i}\right)^2 } 
\end{equation}

The prediction of the new data point is based on the average or majority vote of its neighbor, this method allows the classifier to adapt its prediction according to the local structure of the data which ultimately helps to improve its overall accuracy and flexibility. Since KNN can be used for both classification and regression tasks, its prediction output depends on the type of task (classification or regression). In the case of a classification task, it uses class membership as the output by using the plurality vote of its neighbor to assign the input to the class that is most common among its k nearest neighbors, but when KNN is being used for a regression task, it uses the average of the values of k nearest neighbors as the prediction output, the value of k has an impact on the overall accuracy \cite{assegie2021k} of the model.

\subsubsection{Support Vector Machine (SVM)}
Support Vector Machine (SVM) is a supervised machine algorithm that works by looking for a hyper-plane that creates a boundary between two classes of data to solve classification and regression-related problems \cite{guo2019improving}. It uses the hyper-plane to determine the best decision boundary between different categories in the training dataset, hence they can be applied to vectors that could encode data. Two theories must hold before we can determine the suitability of SVM for certain classification or regression tasks, the first is the availability of high-dimension input space as SVM tries to prevent overfitting by using an overfitting protective measure which is independent of the number of features in the data gives SVM the potential to handle feature spaces in the dataset. The second theory is the presence of linearly separable properties of categorization in the training dataset, and this is because SVM works by finding linear separators between each of the categories to make accurate predictions.

\subsection{Deep Learning Based Classifier}
\subsubsection{Convolutional Neural Network (CNN)}
CNN is a deep learning model with a grid pattern for processing data that is designed to automatically and adaptively learn spatial hierarchies of features, from low- to high-level patterns \cite{yamashita2018convolutional}. It is a mathematical construct that is composed of convolution, pooling, and fully connected layers as three types of layers or building blocks responsible for different tasks for predictions. While convolution and pooling layers, perform feature extraction, the fully connected layer, maps the extracted features into the final output usually known as classification. The convolution layer is composed of mathematical operations (convolution) which plays a very crucial role in Convolutional Neural Networks as in a kind of linear operation. The CNN architecture is a combination of several building blocks like convolution layers, pooling layers, and fully connected layers, and so, a typical architecture consists of repetitions of a stack of many convolution layers and a pooling layer, and then followed by one or more fully connected layers. It stored digital images, and pixel values as a two-dimensional (2D) grid which is an array of numbers along with some parameters called the kernel before an optimizable feature extractor is finally applied at each image position. This makes CNNs a highly efficient classifier for image processing classification tasks, since a feature may occur anywhere in the image. extracted features can hierarchically and progressively become more complex as each layer progressively feeds its output to the next layer, the main task is the minimization of differences between output and ground truth by backward propagation and gradient descent which is an optimization algorithm. This process of optimizing parameters like kernels to minimize the difference between outputs and ground truth is called training.

\subsubsection{Recurrent Neural Network (RNN)}
Recurrent Neural Networks (RNNs) is a type of Neural Network in which output from the previous step is fed to the current step as input, It introduce the concept of memory to neural networks through the addition of the dependency between data points. This addition of dependency between data points ensured that RNNs could be trained to remember concepts by able able to learn repeated patterns. The main difference between RNN and the traditional neural network is the concept of memory in RNN which is made possible as a result of the feedback loop in the cell. Here, it is the feedback loop that enables the possibility of passing information within a layer unlike in feedforward neural networks where information can only be passed between layers. While input and output are independent of each other in a traditional neural network, It is a different ball game in RNN where sequence information is to be remembered, this was made possible in RNN by its Hidden state also known as the memory state through which it remembers previous input to the network, and so it is safe to conclude that the most important features of RNNs is the Hidden state by which it remembers some information in a sequence.
In terms of architecture, RNN architecture is the same as that of other deep neural networks, the main difference lies in how the information flows from the input to the output. While the weight across the network in RNN is the same, deep neural network has different weight matrices for each dense network. The Hidden state in the RNNs which enables them to remember sequence information makes it suitable for natural language processing tasks.

\subsubsection{Long Short-Term Memory (LSTM)}
Long short-term memory (LSTM) network is a recurrent neural network (RNN) that is specifically designed to handle sequential data, such as speech, text, and time series, it is aimed at solving the problem of vanishing gradient in traditional RNNs. It is insensitive to gap length which gives it an advantage over hidden Markov models, hidden Markov models, and other RNNs. It provides a short-term memory for RNN which can last thousands of timesteps thereby making it a "long short-term memory" network. A single LSTM network unit is composed of an output gate, a cell, an input gate, and a forget gate. While the three gates regulate the flow of information into and out of the cell, the cell is responsible for remembering values over arbitrary time intervals as the Forget gates decide on the information to discard from a previous state by assigning a previous state, compared to a current input which assigns a value between 0 and 1. A value of 1 means the information is to be kept, and a value of 0 means the information is to be discarded. The Input gates decide on the exact pieces of new information to store in the current state in the same way as forget gates. Output gates consider both the previous and current states to control which pieces of information in the current state are to output by assigning a value from 0 to 1 to the information. This selective outputting of relevant information from the current state allows the LSTM network to utilize both useful and long-term dependencies in making more accurate predictions in current and future time steps. The fact that they are designed to learn long-term dependencies in sequential data makes them suitable for time series forecasting, speech recognition, and language translation tasks.

\begin{table*}[hbt!]
\centering
\caption{Limitations of current state-of-the-art methods for phishing detection} 
\begin{tabular}{|p{0.7in}|p{0.8in}|p{1.5in}|p{1.5in}|p{1.8in}|}
 \hline
 Author & Dataset  & Research Summary & Method/Algorithm & Limitation  \\
 \hline

 Ann Zeky et al (2023) \cite{magdacy2023detect}  & internally generated dataset  & proposal of extraction based naive Bayes robust model for phishing detection with emphasis on a combination of webpage content and URL feature analysis & naive Bayes \newline URL analyzation \newline webpage content extraction  & 1. problem of bayesian poisoning was not addressed and so the model remains vulnerable to bayesian poisoning     \\ 
\hline

mahdi bahaghighat (2023) \cite{bahaghighat2023high}  & Public URL Dataset & performance comparison of phishing detection method based on several six different algorithm  & naive bayes \newline KNN \newline SVM \newline Random Forest \newline Gradient Boost \newline Logistic Regression & 1. complexity of re-generating tree for every output in random forest remains \newline 2. sole reliance on URL feature attributes means the model remains vulnerable to friendly URL   \\ 
\hline

Nishitha U et al (2023) \cite{nishitha2023phishing}  & Review  & performance comparison of machine learning and deep learning based algorithm for phishing detection  & CNN \newline RNN \newline KNN \newline Random Forest \newline Decision Tree \newline Logistic Regression & 1. 5000 records is too small to train a CNN model and so no confidence here \newline 2. imbalance in the dataset will lead to bias \newline sole reliance on URL feature   \\ 
\hline

Santhosh Raminedi et al (2023) \cite{raminedi2023classification}  & public URL dataset  & evaluation of several machine learning and deep learning based algorithms for phishing detection using URL features  & ANN \newline SVM \newline KNN and Naive Bayes \newline Random Forest \newline Decision Tree \newline Logistic Regression & 1. complexity of generating tree for every output in random forest was not addressed \newline 2. sole reliance on URL feature   \\ 
\hline

Palla Yaswanth and V. Nagaraju (2023) \cite{yaswanth2023prediction}  & phishing dataset  & novel network prediction of phishing sites based on optimal hyper-parameter turning and comparison of the performances of Bayesian and Random Forest Classifier for Phishing Detection  & Naive Bayes and Random Forest & 1. No investigation or hint on the cause of the 5\% failure rate \newline 2. heavy reliance on parameter turning \newline 3. Limited dataset  \\
\hline

 Abdul Karim et al (2023) \cite{karim2023phishing}  & public URL dataset  & proposal of hybrid model with a combination of logistic regression, support vector machine, and decision tree along with a combination of soft and hard voting for efficient defense against phishing attack  & Decision Tree \newline Logistic Regression \newline Support Vector Machine \newline Soft and Hard voting & 1. sole reliance on the URL attribute means the proposed model will be vulnerable to a phishing website with legitimate friendly URL \newline 2. user have to manually surf the internet to get essential URL parameter from a third party to feed the model which is cumbersome and might not be available    \\
\hline

 Ishwarya et al (2023) \cite{ishwarya2023seperation}  & Kaggle email dataset & Proposal of a phishing detection method and performance comparison of Naive Bayes, SVM, KNN, and random forest classifier for phishing detection  & naive bayes \newline KNN \newline SVM \newline Random Forest & 1. use of an imbalance dataset of 87\% ham and 13\% spam which was not addressed \newline 2. Vulnerability to Bayesian poisoning    \\
\hline

 Kamal Omari (2023) \cite{Omari2023}  & UCL Phishing Dataset  & an investigation into the performances and efficiency of Logistic Regression, KNN, SVM, Naive Bayes, Decision Tree, Random Forest, and Gradient Boosting for phishing detection task  & naive bayes \newline KNN \newline SVM \newline Random Forest \newline Gradient Boost \newline Decision Tree \newline Logistic Regression & 1. While Random Forest have a good accuracy of 97.1\%, its complexity of re-generating tree still remains \newline 2. Heavy reliance on URL attributes which means the model remains vulnerable     \\ 
\hline

Twana Mustafa and Murat Karabatak (2023) \cite{mustafa2023feature}  & UCL phishing dataset  & Performance Comparison of different Bayesian classifier based on different Feature Selection Algorithm & naive bayes \newline individual FS \newline forward FS \newline backward FS \newline Plus-I takeaway-r FS \newpage AR1 FS  & 1. each of the Bayesian model remains vulnerable to Bayesian poisoning \newline no hint on why Plus-I takeaway-r FS works better best for naive bayes     \\ 
\hline

 Jaya T et al (2023) \cite{jaya2023appropriate}  & UCL phishing dataset  & usage of frequency weightage of the words for unsupervised clustering of mail into spam and ham messages & naive bayes \newline random forest \newline logistic regression \newline random tree \newline LTSM  & 1. while random forest performed really well, its complexity of generating tree for every output remains a problem \newline possible reason for the poor performance of Bayesian classifier remains to be investigated     \\ 
\hline

\end{tabular}
\end{table*}

\begin{table}[hbt!]
\centering
\caption{Comparison of Different Categories of the State-of-the-art Phishing Detection Methodologies}
\begin{tabular}{|p{0.7in}|p{0.8in}|p{1.5in}| }
 \hline
 Category of Methodologies & Authors & Average Accuracy \\
 \hline

 Naive Bayes-Based & \cite{bahaghighat2023high}, \cite{raminedi2023classification}, \cite{yaswanth2023prediction}, \cite{karim2023phishing}, \cite{Omari2023}, \cite{al2023ofmcdm}, \cite{al2022email}, \cite{ab2022comparative}, \cite{ozker2020content}, \cite{uddin2022comparative}, \cite{rodriguez2019webpages}, \cite{shabudin2020feature}, \cite{sadaf2023phishing}, \cite{alrefaai2022detecting}, \cite{korkmaz2020detection}     & 78.62, 61.0, 95.58, 88.39, 60.1, 95.67, 79.7, 85.15, 83.46, 74.02, 83.88, 92.94, 84.10, 73.8, 70.05     \\
 \hline

  Machine Learning-Based & \cite{bahaghighat2023high}, \cite{raminedi2023classification}, \cite{yaswanth2023prediction}, \cite{karim2023phishing}, \cite{Omari2023}, \cite{alnemari2023detecting}, \cite{al2023ofmcdm}, \cite{aldakheel2023deep}, \cite{rashid2023enhanced}, \cite{arivukarasi2023efficient}, \cite{pandey2023phish}, \cite{sunday2023phishing}, \cite{rugangazi2023detecting}, \cite{karabatak2018performance}        &  95.4, 94.9, 94.6, 78.4, 95.5, 95.7, 96.4, 90.63, 94.7, 94.5, 94.0, 90.0, 97.2, 96.27         \\
 \hline

  Deep Learning-Based & \cite{raminedi2023classification}, \cite{alnemari2023detecting}, \cite{aldakheel2023deep}, \cite{sunday2023phishing}, \cite{huang2019phishing}, \cite{Aljofey_2020}, \cite{adebowale2019deep},  \cite{adebowale2023intelligent}, \cite{huang2019phishing}, \cite{feng2020visualizing}, \cite{somesha2020efficient}, \cite{benavides2022comparative}, \cite{Pratiwi_2018}, \cite{salloum2021phishing}     & 88, 95, 97.4, 93.0, 81.75, 95.02, 92.67, 92.19, 65.9, 99.2, 99.5, 82.0, 83.38, 97.63   \\
 \hline

\end{tabular}
\end{table}

\begin{figure}
\hspace*{-0.6cm}
\centering
\includegraphics[width=1.2\linewidth]{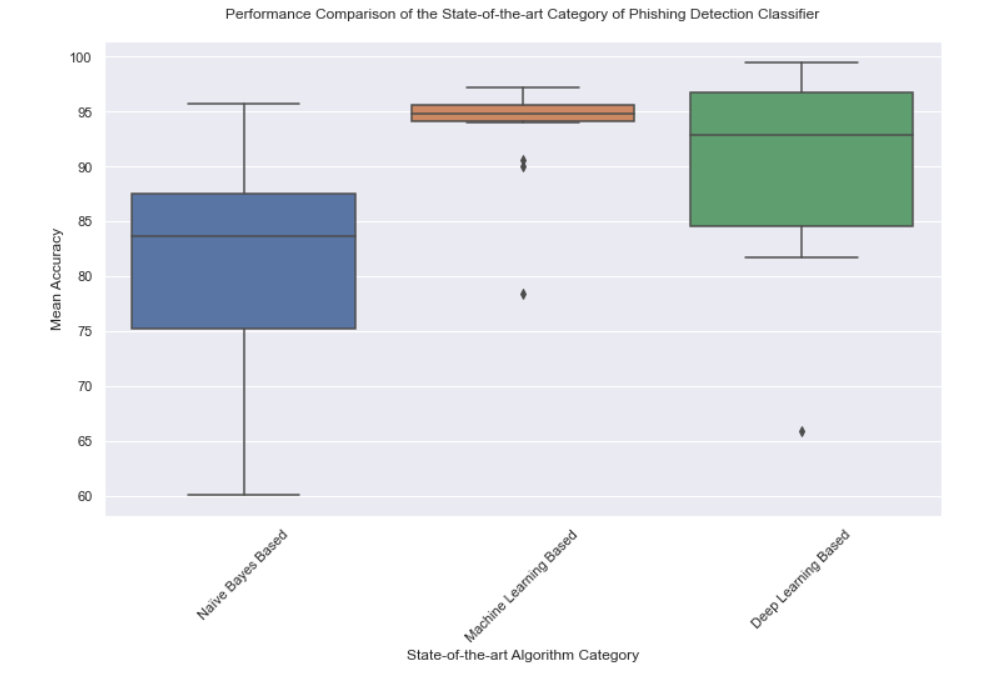}
\caption{\label{fig:Phishing} Comparative Analysis of State-of-the-art Phishing Algorithm for Phishing Detection.}
\end{figure}

\begin{table*}[hbt!]
\centering
\caption{Comparison of Different Algorithms for phishing detection task}
\begin{tabular}{|p{0.6in}|p{0.8in}|p{2.1in}|p{2.3in}|p{0.5in}|}
 \hline
 Classifier & Category & Authors & Accuracy & Average Accuracy\\
 \hline

 Multinomial NB & Naive Bayes  & \cite{bahaghighat2023high}, \cite{raminedi2023classification}, \cite{yaswanth2023prediction}, \cite{karim2023phishing}, \cite{Omari2023}, \cite{al2023ofmcdm}, \cite{al2022email}, \cite{ab2022comparative}, \cite{ozker2020content}, \cite{uddin2022comparative}, \cite{rodriguez2019webpages}, \cite{shabudin2020feature}, \cite{sadaf2023phishing}, \cite{alrefaai2022detecting}, \cite{korkmaz2020detection}       &  78.62, 61.0, 95.58, 88.39, 60.1, 95.67, 79.7, 85.15, 83.46, 74.02, 83.88, 92.94, 84.10, 73.8, 70.05      &  80.431  \\
 \hline

  SVM & Machine Learning & \cite{bahaghighat2023high}, \cite{raminedi2023classification}, \cite{karim2023phishing}, \cite{Omari2023}, \cite{alnemari2023detecting}, \cite{al2023ofmcdm}, \cite{rashid2023enhanced}, \cite{arivukarasi2023efficient}, \cite{pandey2023phish}, \cite{vallepu2023innovative}, \cite{khan2023comparative}, \cite{alrefaai2022detecting}         & 94.43, 94, 71.8, 93.9, 94.0, 94.45, 94.0, 96.4, 95.97, 90.6, 94.0, 59.6     & 
   89.429  \\
 \hline

  Random \newline Forest & Machine Learning & \cite{bahaghighat2023high}, \cite{raminedi2023classification}, \cite{yaswanth2023prediction}, \cite{karim2023phishing}, \cite{Omari2023}, \cite{almseidin2019phishing}, \cite{alnemari2023detecting}, \cite{al2023ofmcdm}, \cite{pandey2023phish}, \cite{aldakheel2023deep}, \cite{rashid2023enhanced}, \cite{rugangazi2023detecting},\cite{khan2023comparative}      & 97.10, 97, 94.6, 96.77, 97.1, 98.11, 97.0, 97.98, 99.13, 94.26, 97.0, 98.6, 97.2        &    97.065  \\
 \hline

  Decision Tree & Machine Learning & \cite{raminedi2023classification}, \cite{karim2023phishing}, \cite{Omari2023}, \cite{alnemari2023detecting}, \cite{al2023ofmcdm}, \cite{rashid2023enhanced}, \cite{pandey2023phish}, \cite{muliono2023phishing}, \cite{khan2023comparative}, \cite{ozker2020content}, \cite{uddin2022comparative}, \cite{alrefaai2022detecting}         & 96.41, 94.9, 96.3, 96.0, 97.02, 93.0, 92.26, 97.62, 95.9, 96.3, 93.57, 93.7        &    95.248  \\
 \hline

  Logistic \newline Regression & Machine Learning & \cite{bahaghighat2023high}, \cite{raminedi2023classification}, \cite{Omari2023}, \cite{arivukarasi2023efficient}, \cite{pandey2023phish}, \cite{sunday2023phishing}, \cite{rugangazi2023detecting}, \cite{khan2023comparative}, \cite{kumar2022machine}, \cite{ab2022comparative}, \cite{ozker2020content}       & 93.16, 92.28, 92.7, 93.4, 92.67, 86.0, 95.9, 96.9, 94.7, 94.18, 86.59    &    92.589   \\
 \hline

  XGBoost & Machine Learning & \cite{bahaghighat2023high}, \cite{karim2023phishing}, \cite{Omari2023}, \cite{al2023ofmcdm}, \cite{ozker2020content}, \cite{uddin2022comparative}, \cite{bahaghighat2023high}, \cite{gualberto2020feature}, \cite{sadaf2023phishing}, \cite{mithra2022website}, \cite{alrefaai2022detecting}     & 96.93, 70.34, 97.2, 96.64, 97.88, 94.79, 99.2, 98.75, 90.83, 96.71, 96.40   &     94.152  \\
 \hline

  KNN & Machine Learning & \cite{bahaghighat2023high}, \cite{raminedi2023classification}, \cite{karim2023phishing}, \cite{Omari2023}, \cite{al2023ofmcdm}, \cite{aldakheel2023deep}, \cite{arivukarasi2023efficient}, \cite{pandey2023phish}, \cite{sunday2023phishing}, \cite{rugangazi2023detecting}, \cite{borra2023k}, \cite{khan2023comparative}, \cite{uddin2022comparative}, \cite{alrefaai2022detecting}           & 95.36, 94.75, 58.63, 95.6, 95.67, 87.0, 93.6, 95.20, 94.0, 97.16, 96.0, 97.2, 83.33, 83.20         &     90.479  \\
 \hline

  CNN & Deep Learning & \cite{huang2019phishing}, \cite{hiransha2018deep}, \cite{Aljofey_2020}, \cite{Zhang_2021}, \cite{adebowale2019deep}, \cite{yerima2020high}, \cite{adebowale2023intelligent}, \cite{somesha2020efficient}, \cite{benavides2022comparative}    & 97.6, 96.8, 95.02, 92.01, 92.55, 98.2, 92.35, 99.43, 84        &     94.218  \\
 \hline

  ANN & Deep Learning & \cite{raminedi2023classification}, \cite{alnemari2023detecting}, \cite{sunday2023phishing}, \cite{mridha2021phishing}, \cite{Pratiwi_2018}, \cite{salloum2021phishing}, \cite{haynes2021lightweight}, \cite{sindhu2020phishing}, \cite{khan2020phishing}, \cite{korkmaz2020detection}      & 88.0, 95.0, 93.0, 98.72, 83.38, 97.63, 97.6, 97.26, 97, 88.22         &     93.581      \\
 \hline

  RNN & Deep Learning & \cite{aldakheel2023deep}, \cite{huang2019phishing}, \cite{adebowale2019deep},  \cite{adebowale2023intelligent}, \cite{halgavs2020catching}, \cite{feng2020visualizing}, \cite{bahnsen2017classifying}, \cite{somesha2020efficient}, \cite{benavides2022comparative}, \cite{soon2020comparison}       & 97.4, 65.9, 92.79, 92.03, 96.74, 99.2, 98.7, 99.57, 80, 93.9      &      91.623      \\
 \hline

\end{tabular}
\end{table*}

\begin{figure}
\centering
\includegraphics[width=1\linewidth]{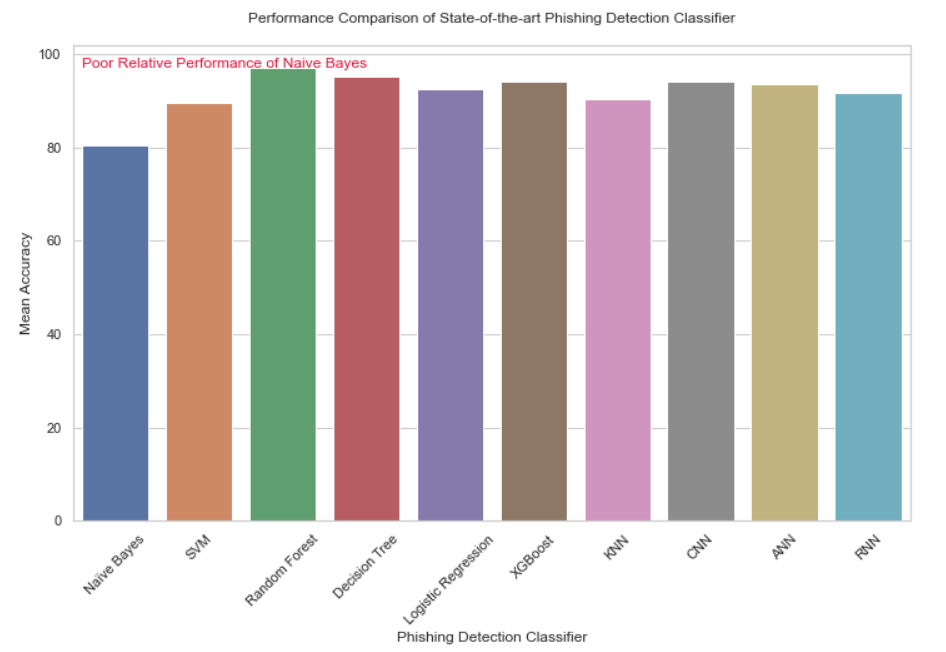}
\caption{\label{fig:Phishing} Comparative Analysis of State-of-the-art Phishing Algorithm for Phishing Detection.}
\end{figure}

\section{Discussion and Analysis}
Our initial decision was to use a combination of f1 score, precision, and accuracy as a performance evaluation metric, but when searching for an appropriate evaluation metric that we could use, we observed that the overwhelming majority of the authors rely solely on accuracy as a measure of evaluation, this shaped our decision to use evaluation as a criterion for performance measurement, and as we all know that there is no single perfect evaluation metric meaning that accuracy alone is not a perfect evaluation metric because different factors and condition can affect the accuracy like imbalance in the dataset which could tilt the accuracy in favor or against a classifier, preprocessing (where roles containing null values were removed or replaces), bias in dataset, possible mistake or negligence on the part of the author and so on. To ensure fairness and a true picture of the performance of individual state-of-the-art algorithms for phishing detection tasks, we decided to use mean accuracy both at individual and categorical levels. 

Having adopted mean accuracy as a measure of performance evaluation to counter the effect of (i) uncertainty in the quality of dataset since they come from a different source in which some are internally generated in certain cases and not available as a public dataset (ii) dataset imbalance or bias that can tilt the result in favor or against a target (iii) series of processing tasks such as complete removal of rows with null values that can cause massive reduction in dataset or replacing them with the mean value which makes the data distorted and not exact (iv)  unintended mistake or negligence as every researcher is different in terms of professionalism, ethical level, attention to details. We tried to look at the reason why phishing is still very effective despite the accuracy and performances of machine learning models, hence, we observed the following;\newline \newline
(1) {Overwhelming reliance on the Uniform Resource Locator(URL) dataset} 

It is worth noting that current state-of-the-art machine-learning phishing detection models are trained based on the properties of the URL such as length of URL, length of the hostname, average words in URL, longest words, character repetition, average path, who is registered domain, domain with copyright, domain age, web traffic, DNS record, google index, PageRank and so on. To better understand why successful phishing attack remains high despite the level of accuracy from state-of-the-art machine learning-based phishing detection model, we will classify the properties of the phishing URL on which ML models are being trained into Controllable Properties and Uncontrollable Properties.
\newline \newline
(a) Controllable Property:\newline 
We classified controllable properties of URLs as properties or characteristics of URLs that could be controlled by attackers.URL characteristics such as length of URL, length of the hostname, average URL, longest word, character repetition, average word, average path, etc can easily be defeated by using SEARCH ENGINE FRIENDLY URL. Except the attacker is not experienced, an experienced attacker will know how the ML model works, and so having a slight experience in web technology will enable a phisher to bypass models that are trained based on the controllable properties of the phishing URL, so the outcome of those models in real-time application after deployment will be an extremely high rate of false negative, and we know that having a high rate of false negative means users will be lead to the phishing site where their credentials will be taken by the attacker.
\newline \newline
(b) Uncontrollable Property:\newline 
We describe them as properties of URLs that cannot be controlled, they build and accumulate over the years. Properties such as domain age, web traffic, Google Index, and PageRank take years to build and accumulate. Hence, for a newly incorporated legitimate business, the website will be relatively fresh and so, properties of their site URL will be very low on these properties, meaning that they will be classified as a phishing website thereby leading to a very high rate of false positive, such ML model are bound to put newly registered legitimate business with quality product and services to offer, startups at a very big disadvantage as their URL will be incorrectly classified or flagged as phishing site.\break

There is a need for significant improvement in training for the existing state of art ML-based model to be truly effective in real-time phishing detection, for this purpose, we proposed combined use of the controllable properties of the URL with web scrapping. Using Uncontrollable properties like the length of the URL, length of the hostname, average URL, longest word, character repetition, average word, average path, etc which takes many years to form will tilt the prediction of the model against recently incorporated legitimate businesses and startups as their site URL will be incorrectly classified as phishing in real-time usage, hence, we suggested the use of the controllable characteristics of site URL along with background web scraping is the background extraction of data from the URL. As attackers have recently resorted to using images on their phishing sites to avoid detection, we are proposing a two-stage prediction model where random forest makes the first prediction based on the properties of the URL, and if the first stage is successful i.e the site is predicted as legitimate, then the model goes to the next stage of prediction where the content of the URL is web scrapped and fed to a Convolutional Neural Network to make the final prediction.

We chose Random Forest for the first stage because it outperformed other machine learning models with a mean accuracy of 97\%  for phishing detection using properties or characteristics of the URL, while we chose CNN due to its effectiveness in natural language processing tasks and image classification. We believe that a two-staged ensemble model consisting of a Random Forest and a Convolutional Neural Network for phishing detection will significantly improve current state-of-the-art phishing detection without jeopardizing the interest of new startups whose domains are relatively new.
. 
\newline \newline

(2) {Overral Poor performance of Naive Bayes at each level} 
\newline
Before any solution can be proposed, it is worth going down memory lane to look at the various assumption of Naive Bayes classifier and its variants, and having affirmed that each Naïve Bayes variants performances and accuracy is largely due to its assumption, Gaussian variant is suitable for anomaly detection due to its assumption that features follow continuous normal distributions, Bernoulli Nave Bayes assumes binomial distribution while multinonial Naive Bayes variant have a dismal performance due to its assumption of discreet multinomial distribution \cite{ige2023performance}, \cite{ige2022ai}, this was also evident during Investigation of the impact of correlation between dataset features on machine learning models for malware classification task \cite{smith2023supervised} where Gaussian Naive Bayes have outlinear performance relative to other classifiers, so each variant of Naive Bayes classifier is parametric based on its individual assumption of feature ditribution in dataset in addition to the generic assumption of independency among feature which rarely holds in real dataset.

The current way of improving the performance of naive Bayesian classifiers is the relaxation of the fundamental assumption of independence among individual attributes in the dataset, which is usually done by an estimation of the joint probability density function (PDF) instead of using the conventional marginal probability density function which is non-naive \cite{wang2013non}. The problem with this approach is that it only gives a slight improvement over the conventional naive Bayes due to the adoption of a joint probability density function, the actual association among the features is not preserved.

We propose regularization to the current Bayes Rule that will put (i) the level of correlation or dependency among the features and (ii) the underlying nature of feature distribution in the dataset into perspective as a way to improve the performance of Naive Bayes-based algorithms. This will invariably lead to a new variant of the Naive Bayes classifier with superior performance compared with the existing variants of the Bayesian classifier.

\section{Conclusion and Future Research Direction}

In this work, we did an extensive review of some of the most recent works on phishing detection, and state-of-the-art algorithms from the past 5 years in order to investigate the performance of Naive Bayes algorithm relative to other state-of-the-art algorithms for phishing detection task, and the factors behind those performances to uncover future research direction. In our comparative study of the performance of Naive Bayes relative to other machine learning and deep learning-based state-of-the-art algorithms for phishing detection tasks through a survey of the recently published research papers, Random Forest, Decision Tree, CNN, XGBoost with an individual mean accuracy of 97.1\%, 95.2\%, 94.2\%, and 94.1\% respectively have the top 4 performance for URL properties-based phishing detection task while Naive Bayes, SVM, RNN with individual mean accuracy of 80.4\%, 89.4\%, and 91.6\% respectively have the worst 3 performance for URL properties-based phishing detection classification task.

In our effort to improve the performance of current state-of-the-art phishing detection methods that rely on the properties of the phishing URLs, especially to counter the ever-evolving phishing methods in which attackers are now using images as text to avoid detection, we proposed a two-stage prediction model where random forest makes the first prediction base on the properties of the URL, and if the first stage is successful i.e the site is predicted as legitimate, then the model goes to the next stage of prediction where the content of the URL is web scrapped and fed to a Convolutional Neural Network to make the final prediction. We chose Random Forest for the first stage because it outperforms other classifiers for phishing detection based on URL properties while CNN was chosen due to its effectiveness in natural language processing and image classification tasks.

Looking at the poor performance of the Naive Bayes classifier both at the individual and categorical levels for which it has the least performance for phishing detection classification task, we propose regularization to the current Bayes Rule that will put both the level of correlation or dependency among the features as well as the underlying nature of feature distribution in a dataset into perspective as a way to improve the performance of Naive Bayes-based algorithms instead of just ignoring them or merely replacing the marginal probability density function with joint probability density function as seen in non-naive Bayes.

\bibliographystyle{plain}
\bibliography{references.bib}
\end{document}